\documentclass[10pt,conference,a4paper]{IEEEtran}

\usepackage{graphicx}
\usepackage[T1]{fontenc}
\usepackage{enumitem}
\usepackage[cmex10]{amsmath}
\usepackage{amssymb}
\usepackage{cite}
\usepackage{amsthm}
\usepackage{textcomp}
\usepackage{siunitx}
\usepackage{color}
\usepackage{subfigure}
\usepackage{cases}
\usepackage[font={small}]{caption}
\usepackage{tcolorbox}
\usepackage{fix-cm}

\makeatletter
\def\thm@space@setup{\thm@preskip=2pt
\thm@postskip=2pt \itshape}
\makeatother
\newtheoremstyle{newstyle}      
{} %Aboveskip 
{} %Below skip
{\mdseries} %Body font e.g.\mdseries,\bfseries,\scshape,\itshape
{} %Indent
{\bfseries} %Head font e.g.\bfseries,\scshape,\itshape
{.} %Punctuation afer theorem header
{ } %Space after theorem header
{} %Heading

\theoremstyle{newstyle}

\newtheorem{theorem}{Theorem}

\theoremstyle{definition}
\newtheorem{example}{Example}
\newtheorem{definition}{Definition}

\theoremstyle{remark}
\newtheorem{remark}{Remark}

\setlist[description]{style=multiline}

\IEEEoverridecommandlockouts

\begin{document}
\sloppy

\setlength{\abovedisplayskip}{0.8mm}
\setlength{\belowdisplayskip}{0.8mm}
\setlength{\abovecaptionskip}{1mm}
\setlength{\belowcaptionskip}{-6pt}

\title{{\fontsize{23}{30}\selectfont Communication-Aware Computing for Edge Processing}}

%\author{
%  \IEEEauthorblockN{Songze~Li}  %\thanks{This research was funded in part by one or all of these grants: %ONR N00014-09-1-0700, CCT-0917343, CCT-1117896, CNS-1213128, ATOSR %TA9550-12-1-0215, and DOT CA-26-7084-00.}
%  \IEEEauthorblockA{
%    University of Southern California\\
%    %songzeli@usc.edu
%    } 
%  \and
%  \IEEEauthorblockN{Mohammad~Ali~Maddah-Ali}
%  \IEEEauthorblockA{Nokia Bell Labs\\
%    %mohammadali.maddah-ali@alcatel-lucent.com
%    }
%  \and
%  \IEEEauthorblockN{A.~Salman~Avestimehr}
%  \IEEEauthorblockA{%Department of Electrical Engineering\\ 
%    University of Southern California\\
%    %avestimehr@ee.usc.edu
%    }
%}

\author{Songze~Li$^{*}$, Mohammad~Ali~Maddah-Ali$^{\dagger}$, and A.~Salman~Avestimehr$^{*}$\\
$^{*}$University of Southern California, $^{\dagger}$Nokia Bell Labs\\
}

\maketitle

\begin{abstract}
We consider a mobile edge computing problem, in which mobile users offload their computation tasks to computing nodes (e.g., base stations) at the network edge. The edge nodes compute the requested functions and communicate the computed results to the users via wireless links. For this problem, we propose a \emph{Universal Coded Edge Computing} (UCEC) scheme for linear functions to simultaneously minimize the load of computation at the edge nodes, and maximize the physical-layer communication efficiency towards the mobile users. In the proposed UCEC scheme, edge nodes create coded inputs of the users, from which they compute coded output results. Then, the edge nodes utilize the computed coded results to create communication messages that zero-force all the interference signals over the air at each user. Specifically, the proposed scheme is universal since the coded computations performed at the edge nodes are oblivious of the channel states during the communication process from the edge nodes to the users. 
\end{abstract}

\vspace{-1mm}
\section{Introduction}
\begin{figure}[htbp]
  \centering
  \includegraphics[width=0.4\textwidth]{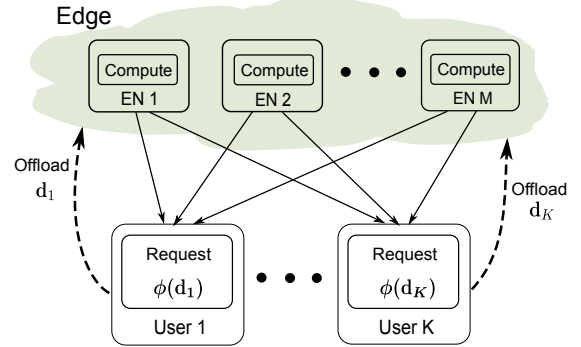}
  \caption{A mobile edge computing system consisting of $K$ mobile users and $M$ edge nodes.}
  \vspace{-5mm}
  \label{fig:setting}
\end{figure}

\vspace{-1.5mm}
We consider a mobile edge computing (or fog computing) scenario (see e.g.,~\cite{beck2014mobile,barbarossa2014communicating,bonomi2012fog}), in which as shown in Fig.~\ref{fig:setting}, a set of $K$ mobile users, denoted by User~$i$ ($i=1,\ldots,K$), offload their computation tasks to a set of $M$ computing nodes scattered at the network edge, which are called the edge nodes and denoted by EN~$j$ ($j=1,\ldots,M$). Each User~$i$ ($i=1,\ldots,K$) has an input ${\bf d}_i$, and requests the computation of an output $\boldsymbol{\phi}({\bf d}_i)$, which is performed at the edge nodes. The overall computation proceeds in two phases: the \emph{computation phase}, and the \emph{communication phase}. In the computation phase, the edge nodes compute all output results for all users. In the communication phase, the edge nodes communicate the the computed results back to the users through wireless links.

One example of the above computing scenario is object recognition and collaborative filtering, which is the key enabler of many augmented reality and machine learning applications. In this case, mobile users are smartphones that want to recognize the pictures captured by their cameras, using a large common database that is stored at the edge nodes (e.g., routers and base stations). During the computation process, the smartphones upload their pictures (or their feature vectors) to the edge nodes. The edge nodes then process the pictures over the database, and return the recognition results to the users. Hence, the pictures (or their feature vectors) correspond to the users' inputs (${\bf d}_i$'s) and the recognition results correspond to the requested outputs ($\boldsymbol{\phi}({\bf d}_i)$'s). Similar operations are also commonly seen in other mobile edge computing applications like navigation and recommendation systems.      

%One example of the above computing scenario is object recognition, which is the key enabler of many augmented reality applications. In this case, mobile users are smartphones that want to recognize the pictures captured by their cameras, using a common database. During the computation process, the smartphones upload their pictures to the edge nodes like routers and base stations that store the database. The edge nodes recognize the pictures over the database, and return the recognition results to the users. Similar operations are also commonly seen in other mobile edge computing applications like navigation and recommender systems.      

We aim to understand how to design the computations across the edge nodes, in order to maximize the physical-layer efficiency (i.e., design the optimum communication-aware computing schemes). We note that as we perform more computations, the spectral efficiency increases. For example, consider the case where we have the same number of edge nodes as the users (i.e., $M=K$), and every single node computes all $K$ outputs (i.e., a computation load of $K^2$). In this case, since all output results are available at all nodes, effectively, we have a $K \times K$ multiple-input single-output (MISO) broadcast channel on the physical layer, which can be diagonalized into $K$ parallel interference-free channels, and the requested outputs of all users can be delivered simultaneously using one unit of communication load.

The above scheme represents one rather trivial point in the set of all possible computation-communication load pairs that accomplish the computation task and deliver the results, denoted by the \emph{computation-communication load region}. In this paper, our objective is to formalize and characterize the entire load region for edge processing. In particular, we focus on \emph{universal} schemes in which the edge nodes perform the computations without knowing the channel gains towards the users in the upcoming communication phase. Universal computation is in fact a common practice in mobile computing systems (see e.g.,~\cite{dinh2013survey,kumar2013survey}), where the computation phase and the communication phase are executed independently of each other. This is primarily due to the fact that the channel state information (CSI) at the communication phase cannot be predicted ahead of time at the computation phase. %Hence, any operation performed in the computation phase should be independent of the future CSI at the communication phase.

%This is an important practical consideration, since the channel state at the communication phase cannot be predicted ahead of time at the computation phase. Hence, any coding performed in the computation phase should be independent of the future channel state at the communication phase.

%computations are often performed without knowing when the communication will take place, and in almost all cases, especially when the channels are time-varying, the channel state at the computation phase will be independent of the channel state at the communication phase. predicting accurate channel state information ahead of time is infeasible. Therefore, motivated by this phenomenon, we ask the following question that is practically relevant:

% In other words, we only consider \emph{universal} edge computing schemes where coding performed in the computation phase is independent of the channel gains. Still, perhaps surprisingly, imposing this universality assumption does not cause any performance loss. That is, there exist universal schemes that achieve the optimum computation-communication load pair for the case that the channel gains are known in the computation phase. Universality allows separate executions of the computation phase and the communication phase, which is a common practice in mobile computing systems where mobile users offload computational-heavy tasks to more resourceful servers and receive computed results (see e.g.,~\cite{dinh2013survey,kumar2013survey})

Our main result is a full characterization of the computation-communication load region  for linear output functions. In particular, we show that the load region is dominated by one corner point that \emph{simultaneously} achieves the \emph{minimum computation load} and the \emph{maximum spectral efficiency}. To establish this result, we first argue that %to save computation and communication loads, 
each edge node should execute \emph{coded} computations, in which computation tasks are executed on some linear combinations of the inputs, rather than executing the task on each input individually. We then propose a \emph{Universal Coded Edge Computing} (UCEC) scheme, in which coded computations are performed at the edge nodes to create messages that neutralize all interference signals in the upcoming communication phase.
In particular, the edge nodes create coded inputs, 
%as many as the total number of original uncoded inputs from the users, 
each as the \emph{sum} of certain carefully selected users' inputs, and pass them into the output functions to compute coded results. 
The coded computation at each node is such that (i) it achieves the minimum computation load of 1 computation unit/user's input, (ii) it is independent of CSI in communication phase, i.e., it is universal, (iii) no matter what the CSI in communication phase is, the coded computation results allow the edge nodes to create messages that  cancel all interference signals over the air and achieve the maximum spectral efficiency of 1 symbol/user/channel use.

%In particular, in the computation phase, the edge nodes create coded inputs, as many as the total number of original uncoded inputs from the users, each as the \emph{sum} of certain carefully selected users' inputs (no channel state information is utilized to perform coding), and pass them into the output functions to compute coded results, achieving the minimum computation load of 1 computation unit/user's input. In the communication phase, utilizing the computed coded results, the edge nodes create messages that cancel all interference signals over the air, achieving the maximum spectral efficiency of 1 symbol/user/channel use.  %Compared with uncoded approaches, the proposed we reduce the computation load by a factor of $K$ while maintaining the maximum spectral efficiency. 

%The key idea to achieve this point is to perform \emph{coded} computations at the edge nodes, such that on average the edge nodes spend one unit of computation on each input of each user, and the edge nodes can utilize these coded results to create messages that neutralize all interference signals at all users. Hence, compared with the above uncoded approach, we reduce the computation load by a factor of $K$ while maintaining the maximum spectral efficiency. 

%that simultaneously minimizes the computation load and the communication load. 

The coding technique in the proposed UCEC scheme is motivated by the ``Aligned Network Diagonalization'' technique in~\cite{shomorony2014degrees}, and the ``Aligned Interference Neutralization'' technique in~\cite{gou2012aligned}, for communications over a two-hop relay network.  In particular, we develop the coded computations, following the patterns of aligned signals recovered at the relays in~\cite{shomorony2014degrees,gou2012aligned}. These aligned signals can be used for signaling over the next hop of the relays such that all interference are cancelled at the destinations. We notice and exploit in this paper the fact that the aligned signals at the relays are independent of CSI of the next hop. While this property is completely irrelevant in communication over the two-hop relay network, it allows us to decouple computation phase from communication phase and form universal computing schemes for edge processing. 

Finally, while in this paper we focus on the case where the entire dataset used to process the requests is stored on each edge node, we can also consider the setting where each edge node only stores a part of the dataset and need to work collaboratively to meet the computational needs. For this setting, recent works~\cite{LMA_all,LMA_ISIT16,li2016fundamental,li2016coded,LMA16_unify,li2017codingfog} have proposed to use coding to minimize the load of communication between edge nodes. An interesting future direction is to design optimal coding schemes for a framework that accounts for communications both from edge nodes to the users and between edge nodes.

\vspace{-2mm}
\section{Problem Formulation}
\vspace{-2mm}
We consider a mobile edge computing problem, in which $K$ mobile users (e.g., smartphones) offload their computation tasks to $M$ edge nodes (e.g., base stations), for some $K,M \in \mathbb{N}$. We denote the $K$ users as User~$1$$,\ldots,$ User~$K$, and the $M$ edge nodes as EN~$1$$,\ldots,$ EN~$M$. User~$k$, $k=1,\ldots,K$, has a sequence of input vectors $({\bf d}_k[i])_{i=1}^{\infty}$ (e.g., pictures of objects in an object recognition application), where for each $i \in \mathbb{N}$, ${\bf d}_k[i] \in \mathbb{R}^{Q}$, for some $Q \in \mathbb{N}$. For each input vector ${\bf d}_k[i]$, User~$k$ wants to compute $B$ output functions (e.g., recognition results) $\phi_1,\ldots,\phi_B: \mathbb{R}^{Q} \rightarrow \mathbb{R}$, for some $B \in \mathbb{N}$. 
%We assume that all computed output functions $\{\phi_1({\bf d}_1[i]),\ldots,\phi_B({\bf d}_1[i]), \ldots, \phi_1({\bf d}_K[i]),\ldots,\phi_B({\bf d}_K[i])\}_{i=1}^{\infty}$ are i.i.d. continuous random variables with mean 0 and variance 1.

% \noindent {\bf Example: Linear Regression.} This type of computation commonly arises in many machine learning algorithms like linear regression. For example, each user wants to train a model she is interested in using a common dataset. Then to compute the gradient based on a dataset matrix ${\bf A} \in \mathbb{R}^{B \times Q}$, in each iteration of the computation, User~$k$ needs to compute the matrix-vector multiplication $[\phi_1({\bf d}_k),\ldots,\phi_B({\bf d}_k)]^T={\bf A}{\bf d}_k$ for the current model ${\bf d}_k$, for all $k=1,\ldots,K$. 
% $\hfill \square$

\noindent {\bf Example: Matrix Multiplication.} One example of the above computation is matrix-vector multiplication, in which each user wants to compute a sequence of output vectors from a dataset matrix ${\bf A}$ stored at the edge nodes. That is, for each input vector ${\bf d}_k[i]$, $i \in \mathbb{N}$, User~$k$ requests the output vector  $[\phi_1({\bf d}_k[i]),\ldots,\phi_B({\bf d}_k[i])]^T={\bf A}{\bf d}_k[i]$, for all $k=1,\ldots,K$. This type of computation commonly arises in many machine learning algorithms. For example, in the gradient decent algorithm for linear regression, when computing the gradient in the current iteration, we need to multiply the data matrix with the model vector from the previous iteration.
$\hfill \square$

In this paper, we focus on \emph{linear} functions such that 
\begin{align}
\vspace{-0.5mm}
\phi_b(\alpha {\bf d}_m[i] + \beta {\bf d}_n[j]) = \alpha \phi_b({\bf d}_m[i]) + \beta \phi_b({\bf d}_n[j]),
\end{align}
for any coefficients $\alpha,\beta \in \mathbb{R}$, $m,n \in \{1,\ldots,K\}$, $i,j \in \mathbb{N}$, and all $b=1,\ldots,B$.

% The computation proceeds over computation blocks of $F$ input vectors at each user, for some $F \in \mathbb{N}$. The operations in the $n$th computation block, $n=1,2,\ldots$, can be decomposed into the \emph{computation phase} and the \emph{communication phase}.  In the computation phase, the $M$ edge nodes compute the output functions for the input vectors ${\bf d}_1[i],\ldots,{\bf d}_K[i]$, for all $i=(n-1)F+1,\ldots,nF$. In the communication phase, the edge nodes communicate the computed results to the intended users. We employ the same computation and communication strategies across blocks, hence in the following detailed description of the computation process we only focus on the first computation block, i.e., $n=1$.

The computation proceeds over a block of $F$ input vectors at each user, for some $F \in \mathbb{N}$. The computation process consists of the \emph{computation phase} and the \emph{communication phase}.  In the computation phase, the $M$ edge nodes compute some output functions from the input vectors ${\bf d}_1[i],\ldots,{\bf d}_K[i]$, for all $i=1,\ldots,F$. In the communication phase, the edge nodes communicate the computed results to the intended users. 

\vspace{-1.8mm}
\subsection{Computation Phase}
\vspace{-1mm}
In the beginning of the computation phase, EN~$m$, $m=1,\ldots,M$, is given $\ell_m$ linear combinations of the users' inputs, denoted by $\mathcal{L}^{(1)}_m,\ldots,\mathcal{L}^{(\ell_m)}_m$, for some $\ell_m \in \mathbb{N}$, i.e., 
\begin{align}
\mathcal{L}^{(j)}_m= \sum_{i=1}^{F}  \sum_{k=1}^{K} \alpha_{mk}^{(j)}[i]{\bf d}_k[i],\label{eq:coded-input}
\end{align}
for some coefficients $\alpha_{mk}^{(j)}[i]\in \mathbb{R}$, and $j =1,\ldots,\ell_m$.

For each linear combination $\mathcal{L}_{m}^{(j)}$ and a subset $\mathcal{W}_m^{(j)} \subseteq \{1,\ldots,B\}$, $j = 1,\ldots,\ell_m$, EN~$m$ computes a function $s_{mb}^{(j)}: \mathbb{R}^Q \rightarrow \mathbb{R}$, such that for each $b \in \mathcal{W}_m^{(j)}$,
\begin{align}
s_{mb}^{(j)} = \phi_b(\mathcal{L}_{m}^{(j)}).
\end{align}

\begin{definition}[Computation Load]
We define the \emph{computation load}, denoted by $r$, as the total number of functions computed across all edge nodes, normalized by the total number of required functions. That is, $r \triangleq \frac{\sum_{m=1}^{M} \sum_{j=1}^{\ell_m} |\mathcal{W}_m^{(j)}|}{FKB}$. $\hfill \Diamond$
\end{definition}

\vspace{-2.3mm}
\subsection{Communication Phase}
\vspace{-1mm}
After the computation phase, the edge nodes communicate the computed results back to the users. We consider a communication scheme that ranges over $T$ time slots, for some $T \in \mathbb{N}$. The symbol communicated by EN~$m$ at time $t$, $t=1,\ldots,T$, denoted by $X_m(t) \in  \mathbb{R}$, is generated as a function, denoted by $\psi_m(t)$, of the functions computed locally at EN~$m$ in the computation phase, for all $m=1,\ldots,M$, i.e.,
\begin{align}
X_m(t) = \psi_m(t)(\{s_{mb}^{(j)}: b \in {\cal W}_m^{(j)}\}_{j=1}^{\ell_m}).
\end{align}
Each edge node has an average power constraint of $P$.

The received symbol at User~$k$ in time $t$, $t=1,\ldots,T$,
\begin{align}
\vspace{-0.5mm}
Y_k(t) = \sum_{m=1}^M h_{km}(t)X_m(t) + Z_k(t),
\vspace{-0.5mm}
\end{align}
where $h_{km}(t) \in \mathbb{R}$, $k=1,\ldots,K$, $m=1,\ldots,M$, is the channel gain from EN~$m$ to User~$k$ in time $t$. The channel gains $\{h_{km}(t): k=1,\ldots,K, \, m=1,\ldots,M\}_{t=1}^{T}$ are time-varying, and they are drawn i.i.d. from a continuous distribution with a bounded second moment. We assume that in the communication phase, %at time $t$, 
the instantaneous channel state information %$\{h_{km}(t): k=1,\ldots,K, \, m=1,\ldots,M\}$ 
is available at all edge nodes. $Z_k(t) \sim \mathcal{N}(0,1)$ is the additive white Gaussian noise at User~$k$ in time $t$.

\begin{definition}[Communication Load]
We define the \emph{communication load}, denoted by $L$, as the total number of communication time slots $T$ normalized by the total number of output functions required by each user, i.e., $L \triangleq \frac{T}{FB}$. $\hfill \Diamond$ 
\end{definition}

After $T$ time slots of communication, User~$k$, $k=1,\ldots,K$, for each $i=1,\ldots,F$, and each $b=1,\ldots,B$, reconstructs the intended output function $\phi_b({\bf d}_k[i])$ using a decoding function $\rho_{kb}[i]$. That is, the reconstructed output function 
\begin{align}
\hat{\phi}_b({\bf d}_k[i]) = \rho_{kb}[i](\{Y_k(t)\}_{t=1}^T).
\end{align}

% We define the distortion of the output function $\phi_b$, $b=1,\ldots,B$, of the input vector ${\bf d}_k[i]$, $i=1,\ldots,F$, of User~$k$, $k=1,\ldots,K$, as

We assume that all input vectors are arbitrary random vectors, and each of the computed output functions is a random variable with finite variance. We define the distortion of the output function $\phi_b$ of the input vector ${\bf d}_k[i]$ from User~$k$ as
\begin{align}
D_{kb}[i] = \mathbb{E}\{(\hat{\phi}_b({\bf d}_k[i])-\phi_b({\bf d}_k[i]))^2\}.
\end{align}

We say that a computation-communicate load pair $(r,L)$ is \emph{achievable}, if there exist a computation scheme with a computation load of $r$ and a communication load of $L$, such that by the end of the communication phase, User~$k$, $k=1,\ldots,K$, can obtain a noisy version of $\phi_b({\bf d}_k[i])$, for all $b=1,\ldots,B$, and all $i=1,\ldots,F$, or more precisely,
\begin{align}
\lim_{P \rightarrow \infty} \frac{\log(1/D_{kb}[i])}{\log P} = 1.
\end{align}

We define the computation-communication load region, denoted by ${\cal C}$, as the closure of the set of all achievable computation-communication load pairs.

\vspace{-2mm}
\section{Motivation and Main Results}
\vspace{-1.5mm}
For the mobile edge computing scenario formulated in the previous section, since each output function needs to be computed at least once, the minimum computation load is at least 1. On the other hand, in the communication phase, even when we can create parallel communication channels for the $K$ users, we need to use the channel at least $FB$ times, one for delivering an output function required by a single user. Hence, the minimum communication load is also at least 1. 

Given the above individual lower bounds on the computation load and the communication load, We ask the following question: \emph{Can we achieve the minimum computation load and the minimum communication load simultaneously? Or is there an edge computing scheme that achieves the load pair $(1,1)$?} %when the number of users equals to the number of edge nodes, i.e., $K=M$, 
We first show through the following example, the achievability of the $(1,1)$ pair when the edge nodes know the channel gains towards the users when executing the computation phase.

% the $(1,1)$ load pair is achievable when the channel state information between the edge nodes and the users in the communication phase is available at the edge nodes in the computation phase, and we illustrate how it is achieved through the following example.

\begin{example}[``Zero-Forcing Ready'' Coded Computing]
We consider a scenario in which $K=2$ mobile users offload their computation tasks to $M=2$ edge nodes. %We consider a matrix-vector multiplication computation task. 
User~$k$, $k=1,2$, has an input vector ${\bf d}_k \in \mathbb{R}^Q$, and wants to compute a length-$B$ output vector ${\bf y}_k = {\bf A} {\bf d}_k$, for some data matrix ${\bf A} \in \mathbb{R}^{B\times Q}$. 

%\vspace{-3.5mm}
\begin{figure}[htbp]
  \centering
  \includegraphics[width=0.48\textwidth]{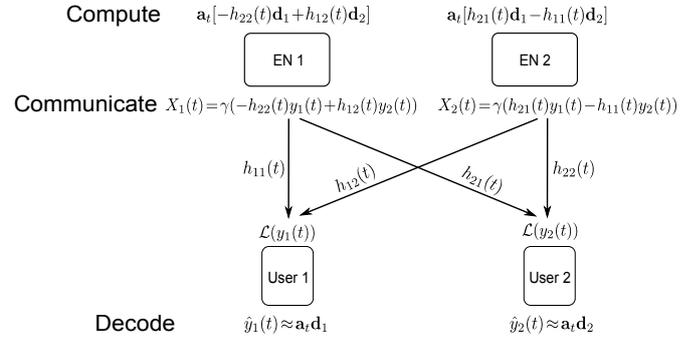}
  \caption{Coded edge computing of $K=2$ users and $M=2$ edge nodes. Using channel state information to design coded computations  allows zero-forcing the interference signal at each user.}
  \vspace{-4mm}
  \label{fig:example-two-EN}
\end{figure}

In this case, we consider a communication phase of $T=B$ time slots, i.e., a communication load of $L=1$, and we assume that the channel gains $\{h_{11}(t),h_{12}(t),h_{21}(t),h_{22}(t)\}_{t=1}^B$ are known at the two edge nodes in the computation phase. 

In the computation phase, for each $t=1,\ldots,B$, EN~1 generates a linear combination of the two input vectors $\mathcal{L}^{(t)}_1 = - h_{22}(t) {\bf d}_1 +  h_{12}(t){\bf d}_2$, and EN~2 also generates a linear combination $\mathcal{L}_2^{(t)} = h_{21}(t) {\bf d}_1 -  h_{11}(t){\bf d}_2$. Then, as shown in Fig.~\ref{fig:example-two-EN}, for the $t$-th row of ${\bf A}$, denoted by ${\bf a}_t$, $t=1,\ldots,B$, EN~1 and EN~2 respectively computes
\begin{align}
s^{(t)}_{1t} &= {\bf a}_t\mathcal{L}^{(t)}_1= -h_{22}(t) y_1(t) +  h_{12}(t)y_2(t),\\
s^{(t)}_{2t} &= {\bf a}_t\mathcal{L}^{(t)}_2= h_{21}(t)y_1(t) -h_{11}(t)y_2(t),
\end{align} 
where $y_k(t)$ is the $t$th element of the output vector ${\bf y}_k$. We perform $B$ coded computations at each of the two ENs, achieving a computation load of $r=1$. 

In the communication phase, at time $t$, $t=1,\ldots,B$, EN~$k$, $k=1,2$, simply sends $X_k(t) =\gamma s^{(t)}_{kt}$, where $\gamma$ is some factor enforcing the power constraint. As a result, as shown in Fig.~\ref{fig:example-two-EN}, User~$k$, $k=1,2$, receives a noisy version of $y_k(t)$, i.e., ${\cal L}(y_{k}(t))$. Therefore, we have successfully performed the computation task, and simultaneously achieved the minimum computation load and the minimum communication load.
% It is easy to see that for a general edge computing scenario with $K$ users and $K$ edge nodes, when the channel state information is available in the computation phase, we can design coded computations to zero-force all interference signals over the air, achieving the optimum load pair $(1,1)$. 
$\hfill \square$
\end{example}

\vspace{-1.5mm}
\begin{remark}
We note that in order to achieve the minimum computation load and the minimum communication load simultaneously, it is critical to perform \emph{coded} computations at the edge nodes, which can create a linear combination of the output functions using one computation unit. Also, the coding needs to be \emph{communication-aware}, such that the computed coded results can be directly utilized to create messages that zero-force the interference signals over the air at each user.  $\hfill \square$
\end{remark}
\vspace{-1.5mm}

While the above example shows the achievability of the optimum load pair $(1,1)$ under the assumption that the channel states are known in prior in the computation phase, this result is not practically interesting. This is due to the fact that the channel states at the communication phase cannot be predicted ahead of time. Hence, %any coding performed in the computation phase should be independent of the future channel states, and 
we should focus on \emph{universal} schemes in which the edge nodes perform the computations without knowing the channel gains towards the users in the future communication phase (i.e., the coefficients in (\ref{eq:coded-input}) are independent of the channel states). Motivated by this phenomenon, we ask the following question:

%This is because in practice computations are often performed without knowing when the communication will take place, and in almost all cases, especially when the channels are time-varying, getting (or predicting) accurate channel state information ahead of time is infeasible. Therefore, motivated by this phenomenon, we ask the following question that is practically relevant:

\begin{tcolorbox}
Is there a \emph{universal} computation scheme that simultaneously achieves the minimum computation load and the minimum communication load, i.e., the load pair $(1,1)$, without requiring channel state information in the computation phase at the edge nodes?
\end{tcolorbox}
\vspace{-1mm}

We answer the above question affirmatively, and present the main result of this paper in the following theorem.

\begin{theorem}
For a mobile edge computing scenario with $K$ mobile users and $K$ edge nodes, there exists a universal computation scheme, named \emph{Universal Coded Edge Computing (UCEC)}, that achieves the minimum computation load and the minimum communication load simultaneously, i.e., the load pair $(1,1)$, for time-varying channels and no channel state information in the computation phase at the edge nodes.
\end{theorem}

% We name the proposed scheme in Theorem~1 that simultaneously achieves the minimum computation load and the minimum communication load as \emph{Universal Coded Edge Computing} (UCEC) scheme, and describe it in Section~IV.
We prove Theorem~1 in Section~V, by describing the proposed UCEC scheme, and analyzing its performance.

\begin{remark}
The key feature of the UCEC scheme is that in the computation phase, without using channel state information, the edge nodes compute coded outputs with a computation load $r=1$. In the communication phase, the edge nodes create messages that admit a communication load $L=1$ and can still neutralize all the interference signals over the air.
$\hfill \square$
\end{remark}
\vspace{-1mm}

\begin{remark}
Theorem~1 implies that when no channel state information is available in the computation phase, the computation-communication load region has a simple shape that is dominated by a single corner point $(1,1)$. Hence, performing computations without being aware of the channel gains does not cause any performance loss. $\hfill \square$
\end{remark}
\vspace{-1mm}

\begin{remark}
In contrast to the ``zero-forcing ready'' scheme in Example~1, we can execute the computation phase separately from the communication phase, without losing any performance. For example, we can perform the computations at some remote edge clusters without knowing when and how the computed results will be delivered to the mobile users, and later have the access points close to the users (e.g., base stations) communicate the results. %and still simultaneously achieve the minimum computation load and the minimum communication load.
$\hfill \square$
\end{remark}
\vspace{-1.5mm}

\begin{remark}
We can directly apply the proposed UCEC scheme in Theorem~1 to the general case of $K$ users and $M$ edge nodes. In particular, when $M > K$, we can use any $K$ out of the $M$ edge nodes to achieve the load pair $(1,1)$. When $M < K$, we can split the $K$ users into $\lceil\frac{K}{M}\rceil$ partitions of size $M$ (except that one partition has size $K-M \lfloor\frac{K}{M}\rfloor$). Then we repeatedly apply the UCEC scheme between the $M$ edge nodes and each of the user partitions, achieving a load pair $(1,\lceil\frac{K}{M}\rceil)$. Overall, the UCEC scheme achieves the load pair $(1,\lceil\frac{K}{M}\rceil)$, for the case of $K$ users and $M$ edge nodes.
$\hfill \square$
\end{remark}
\vspace{-1.5mm}

% \begin{remark}
% When we have more edge nodes than the users, i.e., $M>K$, although the proposed UCEC scheme minimizes the load of computation, in order to minimize the time spent in the computation phase, we also need to pick the $K$ edge nodes with the fastest processing speeds. However, the processing speeds of the edge nodes are often stochastic, and may not be known when the users submit their requests. To combat the randomness of processing speeds and optimize the computation time, in~\cite{LMA_CAECS}, we extend the UCEC scheme to tolerate up to $M-K$ random slow nodes (or stragglers). That is, we can terminate the computation phase and proceed to the communication phase whenever the fastest $K$ out of $M$ edge nodes have finished their local computations, and at the same time, we can still achieve the minimum communication load of $1$.
% $\hfill \square$
% \end{remark}

% In the rest of this section, we illustrate the idea of the proposed UCEC scheme through the following example.

In the next section, we illustrate the key ideas of the proposed UCEC scheme through a simple example.

\vspace{-3mm}
\section{Illustration of the Universal Coded Edge Computing scheme via a simple example}
\vspace{-3mm}

%\begin{example}[Universal Coded Edge Computing] 
% We consider a scenario in which $K=2$ mobile users offload their computation tasks to $M=2$ edge nodes. User~$k$, $k=1,2$, has a block of $F$ input vectors ${\bf d}_k[1],\ldots,{\bf d}_k[F]$, such that ${\bf d}_k[i] \in \mathbb{R}^Q$ for all $i=1,\ldots,F$. For each input vector ${\bf d}_k[i]$, User~$k$ wants to compute a length-$B$ output vector ${\bf a}_k[i] = {\bf A} {\bf d}_k[i]$, for some data matrix ${\bf A} \in \mathbb{R}^{B\times Q}$. In contrast to Example~1, now we do not assume the knowledge of channel gains at the edge nodes in the computation phase. 

We consider a scenario where $K=2$ mobile users offload their tasks to $M=2$ edge nodes. User~$k$, $k=1,2$, has $F=3$ input vectors, and wants to compute the output vectors ${\bf y}_k[i] = {\bf A} {\bf d}_k[i]$, $i=1,2,3$, from some data matrix ${\bf A}$. In contrast to Example~1, now we do not assume the knowledge of channel gains at the edge nodes in the computation phase. 

%We first consider a case $F=3$.
%i.e., User~$k$, $k=1,2$, wants to compute $3$ output vectors ${\bf y}_k[i]={\bf A}{\bf d}_k[i]$, $i=1,\ldots,3$. 

In the computation phase, EN~$1$ generates 
%$2$ linear combinations of the inputs 
${\cal L}_1^{(1)} = {\bf d}_1[1]$ and ${\cal L}_1^{(2)} = {\bf d}_1[2] + {\bf d}_2[1]$,
% \begin{align}
% {\cal L}_1^{(1)} = {\bf d}_1[1], \quad {\cal L}_1^{(2)} = {\bf d}_1[2] + {\bf d}_2[1],
% \end{align}
and EN~$2$ generates ${\cal L}_2^{(1)} = {\bf d}_1[1] + {\bf d}_2[1]$.
% \begin{align}
% {\cal L}_2^{(1)} = {\bf d}_1[1] + {\bf d}_2[1],
% \end{align}
These linear combinations do not depend on the channel gains.

% Then for each $b=1,\ldots,B$, EN~$1$ computes two functions:
% \begin{align}
% s_{1b}^{(1)} &= {\bf a}_b{\cal L}_1^{(1)} = y_{11}(b),\\
% s_{1b}^{(2)} &= {\bf a}_b{\cal L}_1^{(2)} = y_{12}(b) + y_{21}(b),
% \end{align}
% where $y_{ki}(b)$ is the $b$th element of the vector ${\bf y}_k[i]$. Also, EN~2 computes one function 
% \begin{align}
% s_{2b}^{(1)} &= {\bf a}_b{\cal L}_2^{(1)} = y_{11}(b) + y_{21}(b).
% \end{align}

Then for each $b=1,\ldots,B$, EN~$1$ computes two functions $s_{1b}^{(1)} = {\bf a}_b{\cal L}_1^{(1)} = y_{11}(b)$,
$s_{1b}^{(2)} = {\bf a}_b{\cal L}_1^{(2)} = y_{12}(b) + y_{21}(b)$, where $y_{ki}(b)$ is the $b$th element of the vector ${\bf y}_k[i]$. Also, EN~2 computes a function 
$s_{2b}^{(1)} = {\bf a}_b{\cal L}_2^{(1)} = y_{11}(b) + y_{21}(b)$.

In the communication phase, for each $b=1,\ldots,B$, we employ a communication scheme ranging over $2$ time slots. For example, for $b=1$, and some transmit directions ${\bf v}_{11},{\bf v}_{12},{\bf v}_{2} \in \mathbb{R}^2$, we create the transmitted symbols
\begin{align}
\begin{bmatrix} X_1(1) \\X_1(2)\end{bmatrix} &= {\bf v}_{11}s_{11}^{(1)} + {\bf v}_{12}s_{11}^{(2)} \\
&= {\bf v}_{11}y_{11}(1) + {\bf v}_{12}(y_{12}(1) + y_{21}(1)) ,\\
\begin{bmatrix} X_2(1) \\X_2(2)\end{bmatrix} &= {\bf v}_{2}s_{21}^{(1)} = {\bf v}_{2}(y_{11}(1) + y_{21}(1)).
\end{align}

In order to zero-force the interfering signals, we select the transmit directions such that 
${\bf H}_{11}{\bf v}_{12} = -{\bf H}_{12}{\bf v}_{2}$, ${\bf H}_{21}{\bf v}_{11} = -{\bf H}_{22}{\bf v}_{2}$, where ${\bf H}_{km} = \begin{bmatrix} h_{km}(1) & 0 \\ 0 & h_{km}(2)\end{bmatrix}$ is the channel matrix from EN~$m$ to User~$k$ in the two time slots.

After the communication phase, User~1 recovers noisy versions of $y_{11}(1)$ and $y_{12}(1)$ respectively, and User~2 recovers a noisy version of $y_{21}(1)$. Similarly, repeating the same communication process for $B$ times, User~1 can reconstruct ${\bf y}_1[1]$ and ${\bf y}_1[2]$, and User~2 can reconstruct ${\bf y}_2[1]$.

Next, we swap the role of User~$1$ and User~$2$, and perform the same computation and communication operations as before to deliver ${\bf y}_2[2]$ and ${\bf y}_2[3]$ to User~2, and ${\bf y}_1[3]$ User~1.

The above scheme achieves a computation load $r \!=\! \frac{(2+1)\times B \times 2}{3 \times 2 \times B}\!=\!1$, and a communication load of $L\!=\!\frac{2 \times B \times 2}{3 \times B} \!=\!\frac{4}{3}$. Without channel state information, the edge nodes can still exploit coding to reduce the communication load by 33.3\% (the communication load would have been 2 if uncoded computations and orthogonal communications were employed), while maintaining the minimum computation load of $1$. 

The techniques utilized above are motivated by the ``Aligned Interference Neutralization'' (AIN) technique in~\cite{gou2012aligned}, and the ``Aligned Network Diagonalization'' (AND) technique in~\cite{shomorony2014degrees}, for communications over a two-hop relay network. AIN and AND design the transmitted signals at the sources, such that each aligned signal at the relays is the sum of some message symbols, which does not depend on the channel gains on either hop of the network. On the second hop, relays create messages that cancel interference signals over the air at the destinations. 

%In the edge computing problem considered here, in the computation phase, the edge nodes generate coded input vectors, following how the message symbols are aligned at the relays in AIN or AND, which is performed oblivious of the channel gains. As a result, edge nodes can similarly create messages from the coded computation results to zero-force the interference signals over the air at all users.

% We finally note that if in general we consider a block of $F=2W-1$ input vectors, for some $W \in \mathbb{N}$, employing coding techniques motivated by the AIN scheme that is designed specifically for the $2\times2\times2$ relay network, we can use the first $WB$ time slots to simultaneously communicate $WB$ output functions to User~1 and $(W-1)B$ output functions to User~2, and use the second $WB$ time slots to simultaneously communicate $WB$ output functions to User~2 and $(W-1)B$ output functions to User~1. Hence, we can achieve a computation-communication load pair $(1,\frac{2W}{2W-1})$, which goes to the optimal pair $(1,1)$ as $W$ increases.

We finally note that if in general we consider a block of $F=2W-1$ input vectors, for some $W \in \mathbb{N}$, employing coding techniques motivated by the AIN scheme that is designed specifically for the $2\times2\times2$ relay network, we can achieve a computation-communication load pair $(1,\frac{2W}{2W-1})$, which goes to the optimal pair $(1,1)$ as $W$ increases.

%\end{example}

% Having demonstrated the UCEC scheme for the special case of $2$ users and $2$ edge nodes, we describe in the next section, the proposed general UCEC scheme for an mobile edge computing scenario of $K$ users and $K$ edge nodes.

\vspace{-2mm}
\section{Universal Coded Edge Computing Scheme}
\vspace{-1mm}
We prove Theorem~1 by presenting a Universal Coded Edge Computing (UCEC) scheme, for a case of $K$ users and $K$ edge nodes, and time-varying channels. The proposed scheme does not use the channel gains when executing the computation phase, and still asymptotically achieves the optimum computation-communication load pair.

First, we define the set of transmit directions $\Delta_N \triangleq \{0,1,\ldots,N-1\}^{K^2}$, for some arbitrary $N \in \mathbb{N}$. Then, we consider a block of $F = N^{K^2}$ input vectors at each user. For User~$k$, $k=1,\ldots,K$, we assign each of her $N^{K^2}$ input vectors to a unique transmit direction in $\Delta_N$. More specifically, for each element ${\bf p} \in \Delta_N$, we label the input of User~$k$ on the direction ${\bf p}$ as ${\bf d}_{k}^{{\bf p}}$. 

\vspace{-3mm}
\subsection{Computation Phase}
\vspace{-1mm}
At EN~$m$, $m=1,\ldots,K$, for each transmit direction $(p_{11},p_{12},\ldots,p_{KK}) \in \Delta_{N+1}$, we create a coded input vector ${\cal L}_m^{(p_{11},p_{12},\ldots,p_{KK})}$ as the sum of certain input vectors
\begin{align}
{\cal L}_m^{(p_{11},p_{12},\ldots,p_{KK})} = \sum_{k=1}^{K}{\bf d}_{k}^{(p_{11},p_{12},\ldots,p_{km}-1,\ldots,p_{KK})}. \label{eq:input-coding}
\end{align}

Then EN~$m$ computes the functions
\begin{align}
%s_{jb}^{(p_{11},p_{12},\ldots,p_{KK})} &= \phi_b\big(\sum_{k=1}^{K}{\bf d}_{k}^{(p_{11},p_{12},\ldots,p_{kj}-1,\ldots,p_{KK})}\big)\\
s_{mb}^{(p_{11},p_{12},\ldots,p_{KK})} &= \phi_b\big({\cal L}_m^{(p_{11},p_{12},\ldots,p_{KK})}\big)\nonumber\\
& = \sum_{k=1}^{K} \phi_b \big( {\bf d}_{k}^{(p_{11},p_{12},\ldots,p_{km}-1,\ldots,p_{KK})}\big),\label{eq:result-coding}
\end{align}
for all $b=1,\ldots,B$. Here we set $\phi_b({\bf d}_k^{\bf p})=0$ if any element in ${\bf p}$ is $N$ or $-1$. We note that in (\ref{eq:input-coding}), no channel state information is used when creating the coded input vectors. %Also, the coded function $s_{mb}^{\bf p}$ computed at EN~$m$, as shown in (\ref{eq:result-coding}), resembles the linear combination of message symbols decoded at Relay~$m$ in the AND scheme proposed in~\cite{shomorony2014degrees}.

\vspace{-3mm}
\subsection{Communication Phase}
\vspace{-1mm}
In communication phase, the channel state information is known at the edge nodes. At EN~$m$, $m=1,\ldots,K$, for each $b=1,\ldots,B$, we note that the computed functions $\{s_{mb}^{\bf p}: {\bf p} \in \Delta_{N+1}\}$ resemble the aligned signals decoded at the $m$th relay, using the AND scheme in~\cite{shomorony2014degrees} for a $K \times K \times K$ relay network. We perform the communication phase exploiting the communication techniques in AND from the relays to the destinations. Specifically, for channel matrix at time $t$
\begin{align}
{\bf H}(t) =  
\begin{bmatrix}
 h_{11}(t) & \cdots & h_{1K}(t) \\
 %h_{21}(t) & h_{22}(t) & \cdots & h_{2K}(t) \\
 \vdots   & \ddots & \vdots  \\
 h_{K1}(t)  & \cdots & h_{KK}(t) 
\end{bmatrix},
\end{align}
we define 
\begin{align}
{\bf B}(t)=\begin{bmatrix}
 b_{11}(t)  & \cdots & b_{K1}(t) \\
 %b_{12}(t) & b_{22}(t) & \cdots & b_{K2}(t) \\
 \vdots   & \ddots & \vdots  \\
 b_{1K}(t) & \cdots & b_{KK}(t) 
\end{bmatrix} \triangleq {\bf H}(t)^{-1},
\end{align}
and $Q(t)^{\bf p} = Q(t)^{(p_{11},p_{12},\ldots,p_{KK})}\triangleq \prod_{1\leq k,m \leq K}b_{km}(t)^{p_{km}}$.

We demonstrate the communication process to deliver the first function $\phi_1$, and repeat the same process for all other $B-1$ functions. Specifically, we consider a transmission over $d=|\Delta_{N+1}|= (N+1)^{K^2}$ time slots, such that at time $t=1,\ldots,d$, EN~$m$, $m=1,\ldots,K$, communicates a symbol
\vspace{-0.5mm}  
\begin{align*}
X_m(t) \!=\! \gamma \!\!\!\!\! \sum_{{\bf p} \in \Delta_{N+1}} \!\!Q(t)^{\bf p} s_{m1}^{\bf p} \!=\!\! \gamma \!\!\! \sum_{{\bf p} \in \Delta_{N}} \!\! Q(t)^{\bf p}\! \bigg(\sum_{k=1}^K \!b_{km}(t)\phi_1({\bf d}_k^{\bf p})\!\!\bigg),
\end{align*}
where $\gamma$ is some factor to enforce the power constraint.

The received signals at the $K$ users in time $t$, $t=1,\ldots,d$,
\begin{align}
\begin{bmatrix}
Y_1(t) \\
\vdots  \\
Y_K(t)
\end{bmatrix} &= {\bf H}(t)\begin{bmatrix}X_1(t) \\
\vdots  \\
X_K(t) 
\end{bmatrix}+ 
\begin{bmatrix}
Z_1(t) \\
\vdots  \\
Z_K(t)
\end{bmatrix} \nonumber\\
&={\bf B}(t)^{-1} \gamma \!\! \sum_{{\bf p} \in \Delta_{N}} Q(t)^{\bf p} {\bf B}(t) \!\!\begin{bmatrix}
\phi_1({\bf d}_1^{\bf p}) \\ \vdots \\
\phi_1({\bf d}_K^{\bf p})
\end{bmatrix}+
\begin{bmatrix}
Z_1(t) \\
\vdots  \\
Z_K(t)
\end{bmatrix} \nonumber\\
&=\gamma \!\! \sum_{{\bf p} \in \Delta_{N}} Q(t)^{\bf p}\!\!\begin{bmatrix}
\phi_1({\bf d}_1^{\bf p}) \\ \vdots \\
\phi_1({\bf d}_K^{\bf p})
\end{bmatrix}+
\begin{bmatrix}
Z_1(t) \\
\vdots  \\
Z_K(t)
\end{bmatrix}.
\end{align}

As a result, at User~$k$, $k=1,\ldots,K$, all interference signals, i.e., $\{\phi_1({\bf d}_{k'}^{\bf p}): k' \neq k, {\bf p} \in \Delta_{N}\}$, are zero-forced over the air, and the received signals $Y_k(1),\ldots,Y_k(d)$ are $d=(N+1)^{K^2}$ noisy linear combinations of the intended functions $\{\phi_1({\bf d}_k^{\bf p}): {\bf p} \in \Delta_{N}\}$, from which User~$k$ can decode them individually.

% Finally, we repeat the same communication process for all other functions $\phi_2,\ldots,\phi_B$, delivering all computed output functions to the intended users.

Using this scheme, we achieve a computation load $r = \frac{|\Delta_{N+1}|BK}{FKB} = \frac{(N+1)^{K^2}}{N^{K^2}}$,
and a communication load 
$L = \frac{dB}{FB} = \frac{(N+1)^{K^2}}{N^{K^2}}$,
which both go to 1 as $N$ increases.

%\vspace{-1.5mm}
% \section{Conclusions and Future Direction}
% %\vspace{-1.5mm}
% We formulated a ``computation-communication load region'' for mobile edge computing, and proposed  a universal coded edge computing scheme that simultaneously minimizes the computation load and maximizes the spectral efficiency. We re-emphasize two key features of the proposed scheme: 1) edge nodes perform coded computations that are ready for efficient communications, 2) for linear functions, performing coded computations oblivious of the channel states does not cause any performance loss. We view the results of this paper as
%  first steps towards understanding the interplay between computation, communication, and storage in mobile edge computing systems, which is an exciting future research direction.
 %\vspace{-2mm}
 
%\vspace{-1mm}
\section{Acknowledgement}
%\vspace{-1.5mm}
This work is in part supported by NSF grants CCF-1408639, NETS-1419632, ONR award N000141612189, NSA grant, and a research gift from Intel. This material is based upon work supported by Defense Advanced Research Projects Agency (DARPA) under Contract No. HR001117C0053. The views, opinions, and/or findings expressed are those of the author(s) and should not be interpreted as representing the official views or policies of the Department of Defense or the U.S. Government.
\vspace{-1mm}

\bibliographystyle{IEEEtran}
\bibliography{ref-abb}

\end{document}